\documentstyle[prd,aps,eqsecnum,tighten]{revtex}
\begin{document}
% \draft command makes pacs numbers print
\draft
\preprint{JINR E2-97-307, hep-th/9710101}
\title{Direct mode summation for the Casimir energy of a solid ball}
% repeat the \author\address pair as needed
\author{I.~H.~Brevik\thanks{Electronic
address: Iver.H.Brevik@mtf.ntnu.no}}
\address{Division of Applied Mechanics, Norwegian University of
Science and Technology\\ N-7034 Trondheim, Norway}
\author{ V.~V.~Nesterenko\thanks{Electronic address:
nestr@thsun1.jinr.dubna.su} and
I.~G.~Pirozhenko\thanks{Electronic
address: pirozhen@thsun1.jinr.dubna.su}}
\address{Bogoliubov Laboratory of Theoretical Physics, Joint
Institute for
Nuclear Research \\ Dubna, 141980, Russia}
%\date{today}
\maketitle
\begin{abstract}
     The Casimir energy of a solid ball placed in an infinite medium
is calculated by a direct frequency summation using the contour
integration.  It is assumed that the permittivity and permeability of
the ball and medium satisfy the condition $\varepsilon_1
\mu_1=\varepsilon _2\mu_2$.  Upon deriving the general expression for
the Casimir energy, a dilute compact ball is considered
$(\varepsilon_1 -\varepsilon_2)^2/(\varepsilon_1+\varepsilon _2)^2\ll
1$.  In this case the calculations are carried out which are of the
first order in $\xi ^2$ and take account of the five terms in the
Debye expansion of the Bessel functions involved. The implication of
the obtained results to the attempts of explaining the
sonoluminescence via the Casimir effect is shortly discussed.
\end{abstract}

% body of paper here
\section{Introduction}
     The Casimir energy, determined by the first quantum correction
to the  ground  state  of  a  quantum  field  system  with  allowance
for nontrivial  boundary  conditions,  proves  to  be  essential  in
many problems of the elementary particle theory,  in quantum
cosmology, and in physics of condensed  matter.  However,  up  to
now  there  is  no universal  method  for  calculating  the  Casimir
effect for arbitrary boundary  conditions.  This  has  been  done
only  for  simple  field configurations  of  high  symmetry:  gap
between two plates, sphere, cylinder,  wedge and so on.  The
curvature of the boundary and account  of  the dielectric and
magnetic properties of the medium lead to considerable complications.
While the attractive force between two uncharged  metal  plates has
been calculated by Casimir as far back as 1948~\cite{Casimir},  this
effect for perfectly conducting  spherical shell  in vacuum was
computed by Boyer only in 1968~\cite{Boyer} (see also the  latter
calculations \cite{Schw,Balian,LRomeo,NP}).  If  an infinitely  thin
spherical shell  separates  media  with arbitrary dielectric
($\varepsilon_1,  \varepsilon_2$)  and  magnetic  ($\mu_1, \mu_2$)
characteristics, this problem is not solved till
now~\cite{Milton,MiltonNg,MiltonNg2,Carlson}. The main drawback here
is the lack of a  consistent  method  for  removing  the divergences.
Besides an attempt to revive the quasiclassical model of an extended
electron proposed by Casimir~\cite{Casimir2}, interest in this
problem was also initiated by investigations of the bag models in
hadron physics~\cite{Bag,Bag1,Bag2} and recently  by  search  for
the mechanism of sonoluminescence~\cite{sono}.

In this paper we calculate the Casimir energy of a solid ball by
making use of the direct summation of eigenfrequencies of vacuum
electromagnetic     field     by contour integration~\cite{LN,zeta}.
A  definite  advantage  of  this method, compared   with    the
Green's    function    technique    employed
in~\cite{Milton,MiltonNg,MiltonNg2,BrevikK},   is  its  simplicity
and visualization.  We consider a compact  ball placed in an infinite
medium when $\varepsilon_1\mu_1=\varepsilon_2\mu_2$. This condition
enables one to treat the divergencies analogously to the case of a
perfectly conducting spherical shell~\cite{NP}.  Upon deriving the
general expression for the Casimir energy,  we address ourselves to
the case of a dilute  ball $\xi^2\ll
1,\;\xi=(\varepsilon_1-\varepsilon_2)/(\varepsilon_1+\varepsilon_2)$.
The calculations here are of the first order in $\xi ^2$ and take
account of the five terms in the Debye expansion of the Bessel
functions involved.  In  this  way  we  attain  some   generalization
and refinement    of    the    results    obtained    in    this
problem earlier~\cite{BrevikK}.

     The layout of the paper is as follows.  In Sect. II we derive a
general expression for the Casimir energy of  a  solid  ball  in  an
infinite  surrounding  under the condition $\varepsilon_1\mu_1
=\varepsilon_2\mu_2 =c^{-2}$,  where $c$ is an arbitrary constant not
necessary equal to one (it is  the  light velocity    in    the
medium),   the   mode-by-mode   summation   of eigenfrequencies being
used.  In  Sect.  III   the Casimir energy of a dilute  ball
$\xi^2\ll 1$ is calculated.  The implication  of  the obtained result
to the Schwinger attempt to explain the sonoluminescence via  the
Casimir  effect  is  also considered.   In Conclusion (Sect. IV) the
results of the paper are briefly  discussed.  Dispersive effects are
ignored in our paper.

\section{Casimir energy of a solid ball under the condition
$\varepsilon _1\mu_1=\varepsilon_2 \mu_2$}
     Let us  consider  the  Casimir  of  a  solid  ball of radius
$a$, consisting of a material  which is characterized by
permittivity $\varepsilon  _1$ and  permeability  $\mu_1$.  We
assume  that the ball is placed in an infinite medium with
permittivity $\varepsilon  _2$  and  permeability $\mu_2$.  We  also
suppose that the conductivity of the ball material and its
surroundings is equal to zero.

     In our consideration the main part will be  played  by
equations determining  the  eigenfrequencies  $\omega$  of  the
electromagnetic oscillations for this configuration~\cite{Stratton}.
It is convenient to rewrite these equations in terms of the
Riccati-Bessel functions
\begin{equation}
\tilde s_l(x)=x j_l(x),\quad \tilde e_l(x)=xh_l^{(1)}(x),
\end{equation}
where $j_l(x)=\sqrt{\pi/2x}J_{l+1/2}(x)$   is   the  spherical
Bessel function  and  $h^{(1)}_l(x)=\sqrt{\pi/2x}H^{(1)}_{l+1/2}(x)$
is  the spherical  Hankel  function  of  the first kind.  For the
TE-modes the frequency equation reads
\begin{equation}
\Delta^{\text{TE}}_l(a\omega)\equiv \sqrt{\varepsilon_1\mu_2}\,\tilde
s'_l(k_1a)\tilde e_l(k_2a) -\sqrt {\varepsilon_2 \mu_1}\,
\tilde s_l(k_1
a) \tilde e_l '(k_2 a) =0, \label{TE}
\end{equation}
where
$k_i=\sqrt{\varepsilon_i \mu_i}\,\omega, \quad i=1,2$ are the wave
numbers  inside and outside the ball,  respectively;  prime stands
for the differentiation with respect to the argument ($k_1a$ or  $k_2
a$) of  the corresponding Riccati-Bessel function.  The frequencies
of the TM-modes are determined by
\begin{equation}
\Delta^{\text{TM}}_l(a\omega)\equiv \sqrt{\varepsilon_2\mu_1}\,
\tilde s'_l(k_1a)\tilde
e_l(k_2a) -\sqrt {\varepsilon_1 \mu_2}\,\tilde s_l(k_1 a)
\tilde e_l '(k_2 a) =0.
\label{TM}
\end{equation}
The orbital quantum number $l$ in (\ref{TE}) and (\ref{TM}) assumes
the
values $1, 2, \ldots $. Under mutual change $\varepsilon _ i
\leftrightarrow \mu_i, \quad i=1,2$ frequency equations (\ref{TE}) and
(\ref{TM}) transform  into each other.

     It is worth noting that the frequencies of the electromagnetic
oscillations determined by Eq.~(\ref{TE}) and (\ref{TM}) are the same
inside and outside the ball. The physical reason for this is that
photons do not perform  work when passing through the boundary at
$r=a$. This is  in contrast to the case of perfectly conducting
spherical shell in vacuum~\cite{NP}, where eigenfrequencies inside
the shell and outside it are determined by different
equations~\cite{Stratton}.

As usual we define the Casimir energy by the formula
\begin{equation}
E=\frac{1}{2}\sum_s(\omega _s-\bar \omega _s),
\label{energy}
\end{equation}
where $\omega_s$ are the roots of Eqs.~(\ref{TE}) and (\ref{TM}) and
$\bar \omega _s$ are the same roots under condition $a \to \infty $.
Here $s$ is a collective index that stands for a complete set of
indices for the roots of Eqs.~(\ref{TE}) and (\ref{TM}). Denoting the
roots of Eqs.~(\ref{TE}) and (\ref{TM}) by $\omega^{(1)}_{nl}$ and
$\omega^{(2)}_{nl}$, respectively, we can cast Eq.~(\ref{energy}) in
the explicit form
\begin{equation}
E=\frac{1}{2}\sum_{\alpha =1}^2\sum_{l=1}^\infty\sum_{m=-l}^l
\sum_{n=1}^\infty
\left (\omega^{(\alpha)}_{nl}-\bar \omega^{(\alpha)}_{nl}
\right )
= \sum_{l=1}^{\infty}E_l,
\label{explicit}
\end{equation}
where the notation
\begin{equation}
E_l =(l+1/2)\sum _{\alpha =1}^2\sum_{n=1}^\infty\left (
\omega^{(\alpha)}_{nl}-\bar \omega^{(\alpha)}_{nl}
\right )
\label{epartial}
\end{equation}
is introduced. Here we have taken into account that the
eigenfrequencies $\omega ^{(\alpha)}_{nl}$ do not depend on the
azimuthal quantum number $m$. For partial energies $E_l$ we use
representation in terms of the contour integral provided by the
Cauchy theorem~\cite{WW}
\begin{equation}
E_l=\frac{l+1/2}{2\pi i}\oint\limits_C dz\,z\,\frac{d}{dz}\ln
\frac{\Delta_l^{\text{TE}}(az)
\Delta^{\text{TM}}_l(az)}{\Delta^{\text{TE}}_l(\infty)
\Delta^{\text{TM}}_l(\infty)},
\label{Cauchy}
\end{equation}
where the contour $C$ surrounds, counterclockwise, the roots of the
frequency equations in the  right half-plane. Location of the roots
of Eqs.~(\ref{TE}) and (\ref{TM}) enables one to deform the contour
$C$ into a segment of the imaginary axis $(- i \Lambda, i\Lambda)$
and a semicircle of radius $\Lambda $ in right half-plane. At a given
value of $\Lambda $ a finite number of the roots of frequency
equations is taken into account. Thus $\Lambda $ plays the role of a
regularization parameter for the initial  sum in Eq.~(\ref{epartial})
which should be subsequently removed to infinity. In this limit the
contribution of the semicircle of radius $\Lambda$ into
integral~(\ref{Cauchy}) vanishes. From physical considerations it is
clear that multiplier $z$ in (\ref{Cauchy}) is understood to be the
$\lim_{\mu\to 0}\sqrt{z^2+\mu^2}$, where $\mu$ is the photon mass.
Therefore in the integral along the segment $(-i\Lambda, i\Lambda)$
we can integrate once by parts, the nonintegral terms being canceled.
As a result Eq.~(\ref{Cauchy}) acquires the form
\begin{equation}
E_l=\frac{l+1/2}{\pi a}\int\limits_{0}^{\infty}dy
\ln\frac{\Delta^{\text{TE}}_l(iy)
\Delta^{\text{TM}}_l(iy)}{\Delta^{\text{TE}}_l(i\infty)
\Delta^{\text{TM}}_l(i\infty)}.
\label{Cauchy2}
\end{equation}
Now we need the modified Riccati-Bessel functions
\begin{equation}
s_l(x) =\sqrt{\frac{\pi x}{2}}I_\nu (x), \qquad e_l(x)=\sqrt
{\frac{2 x}{\pi}}K_\nu (x), \quad \nu=l+{1}/{2},
\end{equation}
where $I_\nu(x)$  and $K_\nu(x)$ are the modified Bessel
functions~\cite{AS}.
With allowance for the asymptotics of $s_l(x)$ and $e_l(x)$
at $x\to\infty$ and fixed $l$
\begin{eqnarray}
s_l(x)&\simeq& \frac{1}{2}e^x, \\
e_l(x) &\simeq & e^{-x}
\end{eqnarray}
equation~(\ref{Cauchy2}) can be rewritten as
\begin{eqnarray}
\label{general}
E_l&=&\frac{l+1/2}{\pi a}\int \limits_0^\infty dy \ln \left \{
\frac{4 e^{-2(q_1-q_2)}}{(\sqrt{\varepsilon_1 \mu_2}+
\sqrt{\varepsilon _2\mu_1})^2}
\right . \nonumber \\
&&\times  \left [\sqrt{\varepsilon _1\varepsilon _2\mu_1\mu_2}\left (
(s'_l(q_1)e_l(q_2))^2+(s_l(q_1)e_l'(q_2))^2
\right )
\right .      \\
&&-
(\varepsilon _1 \mu_2 +\varepsilon _2\mu_1)
s_l(q_1)s'_l(q_1)e_l(q_2)e_l'(q_2)\bigr ]
\biggr\},\nonumber
\end{eqnarray}
where $q_i=\sqrt{\varepsilon _i\mu_i}\,y,\quad i=1,2$. We shall use
this general equation in the next Section but here  we address
ourselves to the special case when the condition
\begin{equation}
\varepsilon _1 \mu_1=\varepsilon _2\mu_2=c^{-2} \label{condition}
\end{equation}
is fulfilled. Here $c$ is an arbitrary positive
constant (the light velocity in medium). Physical implications
of this condition
at $c=1$ can be found in~\cite{Lee}. Now Eq.~(\ref{general}) is
simplified considerably
\begin{eqnarray} \label{special}
E_l&=&\frac{c(l+1/2)}{\pi a}\int \limits_{0}^{\infty}dy \ln \biggl \{
\frac{4}{\varepsilon +\varepsilon ^{-1}+2}\Bigl [ (s'_l(y) e_l(y))^2+
(s_l(y) e'_l(y))^2 \nonumber \\
&& - (\varepsilon +\varepsilon ^{-1})s_l(y) s'_l(y) e_l(y) e'_l(y)
\Bigr ] \biggr \},
\end{eqnarray}
where $\varepsilon =\varepsilon _1/\varepsilon _2$. The argument
of the logarithm in
(\ref{special}) can  be transformed, if the following two
equalities for the functions
$s_l(y)$ and $e_l(y)$
\begin{eqnarray}
s'_l(y) e_l(y) -s_l(y) e'_l(y)& =& 1 , \label{eq1}\\
s'_l(y) e_l(y) +s_l(y) e'_l(y)& = & (s_l(y) e_l(y))'. \label{eq2}
\end{eqnarray}
are taken into account. It gives
\begin{equation}
E_l=\frac{c(l+1/2)}{\pi a}\int\limits_{0}^{\infty}dy \ln \left \{
1-\xi ^2\left [(s_l(y) e_l(y))'\right ]^2
\right \},
\label{final}
\end{equation}
where
\begin{equation}
\xi =\frac{\mu_2 -\mu_1 }{\mu_2+\mu_1}=\frac{\varepsilon _1
-\varepsilon _2}
{\varepsilon_1+\varepsilon_2},\quad \varepsilon _i\mu_i=c^{-2},\;
i=1,2.
\end{equation}
Thus, for a ball with a vacuum on the outside, $\xi=(1-\mu)/(1+\mu)=
(\varepsilon -1)/(\varepsilon +1)$ and $c=1$.  The expression
(\ref{final}) agrees with the results obtained in~\cite{MiltonNg,ML},
if one performs a partial integration of the expression for $E$ given
in these references and puts the cutoff parameter $\delta$ equal to
zero.  If $\varepsilon =0$ or $\infty$ and $c=1$ then, as one could
expect, Eq.~(\ref{final}) turns into the analogous expression for
the perfectly conducting spherical shell in vacuum~\cite{Schw,NP}.

  We remark that in a previous paper~~\cite[(1982)]{BrevikK} an
expression for the Casimir energy was calculated that seemingly is in
conflict with Eq.~(\ref{final}). Namely, Eq.~(2.42) in that paper
corresponds to the following expression for~$E_l$, assuming
$\mu_1=\mu,\;\mu_2 =1$ as above, and putting the cutoff parameter
equal to zero,
\begin{equation}
E_l=- \frac{(\mu -1)^2}{\pi a }\nu \int^\infty _0 dx
\frac{s_ls_l\,'e_le_l'}{D_l\tilde D_l}\,x\, \frac{d}{dx}\ln (1-\lambda _l^2).
\label{new}
\end{equation}
Here $\lambda _l=(s_le_l)'$, and $D_l,\;\tilde D_l$ are defined by
\begin{eqnarray}
D_l(x)=\mu s_l(x)e_l'(x)- s'_l(x)e_l(x),\\
\tilde D_l(x)=\mu s_l'(x)e_l(x)- s_l(x)e_l'(x).
\end{eqnarray}
It turns out, however, that these two equations (\ref{final}) and
(\ref{new}) are in agreement.   To show the equivalence is not quite
trivial, but follows after some algebra taking into account the
derivatives of the logarithms and the Wronskian~(\ref{eq1}). we omit
the details here. The expression (\ref{final}) is the most convenient
form to work with, and we will use it in the following where  we turn
to the special case of a dilute medium satisfying the condition
$\varepsilon \mu=1$.  We emphasize that the expression (\ref{final})
is general, making no restriction at all on the magnitude of
$\mu_1/\mu_2$.

\section{Casimir energy of a dilute compact ball}
Now we address  ourselves to the consideration of the Casimir energy
of a dilute compact ball when
\begin{equation}
|\xi |\ll 1.
\end{equation}
That means, to the lowest order in~$\xi$,
\begin{equation}
\ln (1-\xi ^2 \lambda _l^2) \simeq -\xi ^2 \lambda ^2_l,
\end{equation}
which reflects a general property of all Casimir calculations in
dilute media: the lowest order correction for all physical quantities
is proportional to the {\it square} of the susceptibility (electric
or magnetic).  We shall henceforth work only to the second order
in~$\xi$. From (\ref{final}) we then get for the Casimir energy
\begin{equation}
E=- \frac{\xi ^2}{\pi a}\sum_{l=1}^{\infty}\nu \int_{0}^{\infty} dx\,
\lambda_l^2(x).
\label{E}
\end{equation}
For simplicity we have put here the constant $c$ in condition
(\ref{condition})  to be equal~1.

We now invoke the following useful expansion,which was worked out by
one of us some time ago~\cite{Breviknew}
\begin{eqnarray}
\lambda_l &=&\frac{t^3}{2\nu}\left [
1-\frac{1}{8 \nu^2}(2-27\, t^2+60\, t^4 -35\, t^6)  \right .
\nonumber \\
&&{}- \frac{1}{128 \nu^4}(108\, t^2 -3615\, t^4+21420\, t^6 -47250\,
t^8  \nonumber \\
&&{} +44352\, t^{10}-
15015\, t^{12})+ {\cal O}(1/\nu^6)
\biggr ].
\label{lambda}
\end{eqnarray}
Here, $t(z)=(1+z^2)^{-1/2},\; z=x/\nu$. This expression is based upon
the Debye expansions for the modified Bessel functions~\cite{AS}. From
(\ref{lambda}) we calculate
\begin{eqnarray}
\lambda ^2_l&=& \frac{t^6}{4\nu ^2}\left [1-\frac{1}{4\nu^2}(2-27\,t^2+
60\,t^4-35\,t^6)\right . \nonumber \\
&&{}+\frac{1}{16\nu^4} (1-54\,t^2+1146\,t^4-6200\,t^6+13185\,t^8
\nonumber \\
&&{}-12138\,t^{10}+4060\,t^{12})+ {\cal O}(1/\nu^6)\biggr ],
\end{eqnarray}
which can now be inserted into Eq.~(\ref{E}). From the integral
representation of the beta function~\cite{AS} ${\rm B}(q,p)=\Gamma
(q)\Gamma (p)/\Gamma (q+p)$ we derive the formula
\begin{equation}
\int_{0}^{\infty}t^p(z)\, dz =\frac{\sqrt \pi}{2}\,\frac{\Gamma
 \left (\displaystyle\frac{p-1}{2}\right )}
{\Gamma\left (\displaystyle \frac{p}{2}\right )}\,{,}
\end{equation}
which is quite useful in the present context. After some calculation
we obtain
\begin{equation}\label{Esums}
E=- \frac{3\xi ^2}{64 a} \left [\sum_{l=1}^{\infty } \nu^0
-\frac{9}{128}
\sum_{l=1}^{\infty}\frac{1}{\nu ^2} +\frac{423}{16384}
\sum_{l=1}^{\infty}
\frac{1}{\nu^4}+ {\cal O}\left (
\frac{1}{\nu ^6}\right )
\right ]{.}
\end{equation}
Here, the first sum will be regularized by means of the Riemann zeta
function~\cite{Obook}, which turns out to be most useful in all
Casimir problems involving nondisperssive media. In practical
calculations the only formula needed is
\begin{equation}
\sum_{l=0}^{\infty}\nu^s=(2^{-s}-1)\zeta (-s)\,{,}
\end{equation}
>from which $\sum_{l=1}^{\infty}\nu ^0=-1$. Moreover, since the two
last sums in (\ref{Esums}) are known,
\begin{equation}
\sum_{l=1}^{\infty}\frac{1}{\nu^2}=\frac{1}{2}\pi^2-4,\quad
\sum_{l=1}^{\infty}
\frac{1}{\nu^4}=\frac{1}{6}\pi ^4-16\,{,}
\end{equation}
we get, omitting the remainder in~(\ref{Esums}),
\begin{equation}
E=\frac{3\xi ^2}{64 a}\left [1+\frac{9}{128}\left  (\frac{1}{2}
\pi ^2-4 \right )
-\frac{423}{16384}\left ( \frac{1}{6}\pi^4 -16
\right )
\right ].
\label{Efinal}
\end{equation}
The energy is positive, corresponding to a {\it repulsive} surface
force. Remember, though, that we are working here with the {\it
nondispersive} theory only.

The structure of the three different terms in (\ref{Efinal}) is the
following. The first term stems from the order $1/\nu$ in the Debye
expansion. This term agrees with the results obtained by
Milton~\cite{Milton} and Milton and Ng~\cite{MiltonNg} in the case
when the condition~(\ref{condition}) holds. Numerically, the three
terms between square brackets in (\ref{Efinal}) are
$[1+0.06573-0.00610]$. Thus the second term, stemming from the order
$1/\nu^3$ in the Debye expansion, describes a repulsive correction of
about 6.6 per cent. Finally the third term, stemming from the order
$1/\nu^5$ in the Debye expansion, describes a 0.6 per cent attractive
correction. We have thus improved the calculations
in~\cite{Milton,MiltonNg} by four orders in magnitude. The next
correction, not included here, is of order $1/\nu^7$ in the Debye
expansion.

 Strictly speaking our consideration is applicable only when the
condition $\varepsilon_i\mu_i=1$ is satisfied.  However keeping in
mind a smooth dependence of the Casimir energy on the parameters
specifying the field configuration, one can expect that the final
formula can also be used in the case of a nonmagnetic dilute
dielectric ball placed in vacuum (or a spherical cavity in an
infinite dielectric surrounding) at least for estimation only.
Therefore we present here some numerical estimations relating to
sonoluminescence.  With a good accuracy one can put in dimensional
units
\begin{equation}
E\simeq 3 \hbar c\xi ^2/(64 a).
\label{dim}
\end{equation}
Take, as an example, $|\xi|=0.1, \; a=4\cdot10^{-4}$~cm. Then
$E\simeq 2\cdot 10^{-5}$~eV. This is immensely  smaller than the
amount of energy ($\sim 10$~MeV) emitted in a sonoluminescent flash.
Furthermore, the Casimir energy (\ref{dim}), being positive,
increases when the radius of the ball decreases. The latter
eliminates completely the possibility of explaining, via the Casimir
effect, the sonoluminescence  for bubbles in a liquid.  As
known~\cite{PT}  emission of light takes place at the end of the
bubble collapse. Recently an important experimental studies have been
done to measure the duration of the sonoluminescence
flash~\cite{last}. In view of all this it is difficult to imagine
that the Casimir effect, at least in its nondispersive version,
should be important for the sonoluminescence  phenomenon.

\section{Conclusion}
Our method for calculating the Casimir energy $E$ by means of the
contour integral (\ref{Cauchy}) proves to be very convenient and
effective. As known, there are in principle at least two different
methods for calculating $E$: one can follow a local approach,
implying use of the Green's function  to find the energy density (or
the surface force density). Or, one can sum the eigenfrequencies
directly. Equation (\ref{Cauchy}) thus means that we have adopted the
latter method here.  The Cauchy integral formula  turns out to be
most useful in other context also, such as in the calculation of the
Casimir energy for a piece wise uniform relativistic
string~\cite{BE}.  A survey  on this subject can be found
in~\cite{BBO}. The great advantage of the method is that the
multiplicity of zeros in the dispersion function is automatically
taken care of, i.e., one does not  have to plug in the degeneracy in
the formalism by hand.

A remarkable feature of our approach is also the ultimate formula for
the Casimir energy having the form of the spectral representation,
i.e., of an integral  with respect to frequency between the limits
$(0,\infty)$ of a smooth function, spectral density. Evidently, for
physical applications  one needs to know the frequency range which
gives the main contribution into the spectral density. An example of
this representation for the partial energies $E_l$ is
Eq.~(\ref{final}), where the substitution $y=\omega a$ should be
done. As shown above, the partial energies $E_l$ decrease rapidly as
$l$ increases. Therefore the most interesting is a few first values
of $l$.  In this case, as one could expect,  the spectral density is
different from zero when $\omega a \simeq 1$.  Keeping in mind the
search for the origin of the sonoluminescence we
put~\cite{MiltonNg,PT} $a=4\cdot 10^{-4}$~cm. Then the wave length of
the photon in question turns out to be $25.0\cdot 10^{-4}$~cm, i.e.,
this radiation belongs to infrared region, while  in experiments on
sonoluminescence the blue light is observed~\cite{PT}.  This fact
also testifies against the possibility of explaining the
sonoluminescence by the Casimir effect.

It is worth noting that  the spectral distribution of the Casimir
energy is practically not discussed in literature   while the space
density of this energy has been investigated in detail (see, for
example~\cite[(1983)]{BrevikK}. From the physical point of view the
space density and spectral density of energy in this problem should
be treated on the same footing. One can remind here the treatment of
the Casimir effect as a manifestation of the fluctuations of the
vacuum fields~\cite{LL}, these fluctuations being occurred in space
and time simultaneously.

It should be emphasized that in this paper we have neglected the
dispersion effects when calculating the  Casimir energy. Importance
of this point has been demonstrated in~\cite{BSS}.  As for the
elucidation of the sonoluminescence origin, we have to stress once
more that in our consideration we have contented ourselves with the
static Casimir effect only.

\acknowledgments

This work was accomplished with financial support of Russian
Foundation of Fundamental Research (grant ¤ 97-01-00745).

\end{document}